# *In situ* Auger electron spectroscopy of complex oxide surfaces grown by pulsed laser deposition

Thomas Orvis, Mythili Surendran, Yang Liu, Austin Cunniff

Mork Family Department of Chemical Engineering and Materials Science, University of Southern California, Los Angeles, 90089

Jayakanth Ravichandran[a]

Mork Family Department of Chemical Engineering and Materials Science, University of Southern California, Los Angeles, 90089

Ming Hsieh Department of Electrical Engineering, University of Southern California, Los Angeles, 90089

[a] Electronic mail: jayakanr@usc.edu

The authors report *in situ* Auger electron spectroscopy (AES) of the surfaces of complex oxides thin films grown by pulsed laser deposition (PLD). The authors demonstrate the utility of the technique in studying chemical composition by collecting characteristic Auger spectra of elements from samples such as complex oxide thin films and single crystals as well as metal foils. In the case of thin films, AES studies can be performed with single unit cell precision by monitoring thickness during deposition with reflection high energy electron diffraction (RHEED). The authors address some of the challenges in achieving *in situ* and real time AES studies on complex oxide thin films grown by PLD. Sustained layer-by-layer PLD growth of a $CaTiO_3$/$LaMnO_3$ superlattice allows depth-resolved chemical composition analysis during the growth process. The evolution of the Auger spectra of the elements from individual layers were used to perform chemical analysis with monolayer-depth resolution.

# I. INTRODUCTION

Complex oxide materials demonstrate unconventional physical properties such as ferroelectricity,[1] sharp metal-insulator transitions,[2] colossal magnetoresistance,[3] and high-temperature superconductivity.[4] Utilization of these properties to create modern devices, such as thin-film transistors and resistive random-access memory,[5,6] has motivated interest in the low-dimensional properties of these materials. The deposition of complex oxides with atomic-layer precision can be accomplished with various techniques, but a common method especially suited to complex oxides is pulsed laser deposition (PLD).[7-10] However, the growth mechanisms of PLD are notoriously complex with compositional (and property) sensitivity to growth conditions,[11-13] which is a great impediment to achieving reliable properties due to stoichiometry dependence,[14-17] or if the system is sensitive to interfacial termination.[18,19] As such, *in situ* compositional analysis of complex oxide thin film surfaces during deposition has the potential to identify and more accurately address stoichiometry errors arising from process parameters, which limit precise control of physical properties.

Reflection high energy electron diffraction (RHEED) has been used as an *in situ* technique for monitoring surface structural quality and reconstructions in a molecular beam epitaxy system since as early as 1969.[20] However, it was not until 1981 that observation of intensity oscillations during deposition was correlated to growth rate.[21] With the advent of PLD as a technique for the growth of complex oxide thin films,[22] differential pumping systems for the electron sources were implemented to accommodate the higher growth pressures.[23,24] This development led to a dramatic improvement in the understanding of the growth mechanisms during PLD, and was fundamental to the advancement of atomic layer engineering of complex oxides.[25] However, deriving compositional information from the structural data provided by RHEED is indirect and qualitative, and compositional evolution during PLD growth remains relatively underexplored.

Auger electron spectroscopy (AES) has been utilized as a powerful surface composition analysis technique for over fifty years.[26] Yet for more than half that period AES was an *ex situ* technique, typically used for identifying surface contamination.[27] Realizing *in situ* AES required considerable design efforts even for ultra-high vacuum growth methods such as molecular beam epitaxy (MBE) or atomic layer deposition,[28,29] and required *in vacuo* chamber transfer or low pressure deposition for PLD systems dedicated to growing non-oxide materials such as metals or carbides.[30,31] However, recent developments in probe design by Staib Instruments now allow real-time *in situ* observation of surface composition during deposition without inhibiting the growth process in systems which rely on a higher background pressure such as oxide PLD.[32-34] This has been achieved by creating a probe, which uses an electrostatic lens system coupled to a retarding field analyzer (RFA), in contrast to cylindrical mirror analyzers (CMAs) more commonly used for AES. Although RFAs have been used for AES for decades, they have historically suffered from higher noise than their CMA counterparts.[26] However, this probe has implemented a collimator lens designed such that the measured signal is proportional to energy, thereby increasing its sensitivity and yielding much improved signal-to-background and signal-to-noise. This, in turn, allows the probe to operate under harsh deposition conditions with a large working distance while providing fast acquisition times, stability against long-term material deposition, and a wide range of operating pressures. The large working distance decreases the effective solid angle but prevents interference with the deposition source. Fast acquisition times are necessary for analysis on a scale shorter than the time needed to deposit a single monolayer of material. Figure 1 shows the *in situ* AES probe incorporated into our PLD system with *in situ* RHEED, where both the techniques share the same electron gun source. The robustness of this probe offers a unique opportunity to address a wide range of questions regarding the compositional kinetics of complex oxide thin film growth using PLD.

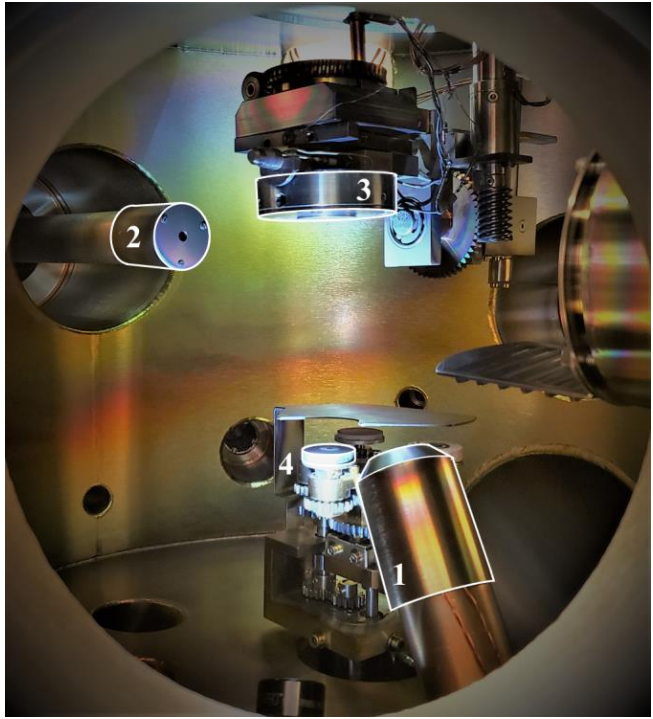

FIG. 1. Photograph of the inside of our PLD chamber with Auger probe (1), electron gun (2), substrate heater (3), and target (4). Note that the Auger probe is positioned at 100 mm from the substrate in this picture, as opposed to the nominal working distance of 55 mm.

In this work, we report *in situ* Auger spectra of numerous complex oxide materials, and several metals, in our PLD growth chamber, demonstrating the versatility of this probe. The viability of the probe for high-pressure acquisitions is also shown by studying the evolution of Auger signal from a $NdGaO_3$ substrate as a function of background oxygen partial pressure. Finally, we demonstrate the use of the *in situ* AES probe during the pulsed laser deposition of a $CaTiO_3$/$LaMnO_3$ (CTO/LMO) thin film superlattice on a $NdGaO_3$ substrate at various thickness intervals to identify the surface composition. Layer-by-layer growth is demonstrated by reflection high energy electron diffraction (RHEED) and *in situ* AES is used to observe the compositional changes during the growth. We determine the depth sensitivity of the AES using these results.

## II. METHODS

### A. Auger Electron Spectra Acquisition

Auger electron spectra were collected with a Staib Instruments AugerProbe™. The excitation was provided by a Staib electron gun operating in the grazing incidence geometry at an excitation voltage of 5 kV and an emission current of 5 μA. The probe was positioned at a working distance of 55 mm 30° off the substrate normal, giving an approximate acceptance solid angle of 0.4% of $2\pi$.[32] The probe signal was optimized, as per the manufacturer's instructions, prior to every sustained spectra acquisition by adjusting the position of the electron beam on the sample, the tilt angle of the sample relative to the probe, and the X and Y components of the probe's built-in magnetic steering. After maximizing the signal, individual spectra acquisitions can be optimized by adjusting the photomultiplier voltage, signal modulation, energy resolution, time constant, and acquisition time. To preserve the integrity of comparative scans the adjustable parameters of spectrum acquisition were not changed between successive acquisitions.

During the CTO/LMO superlattice growths, spectra were taken at specific intervals of deposited layers: after 1, 2, 3, 4, 5, 7, 10, 15, and 20 monolayers of CTO and LMO during the first deposition of each material, and after 2, 5, 10, and 20 monolayers of each material during the second deposition. During the acquisition of Auger spectra, the chamber pressure was reduced to $< 10^{-6}$ mbar and the accelerating voltage of the electron source decreased from the 35 kV used for RHEED during deposition to 5 kV. The tilt of the film relative to the electron beam was increased from ~1° to 15° to maximize the Auger signal. Tilt angle was controlled with a computerized microcontroller, thereby allowing translation between the same precise tilt angles for deposition and AES. The spectra were collected using Staib Instruments' proprietary LabVIEW-based software. After acquisitions the state of the chamber was returned to that used for deposition, and vice-versa.

### B. Data Processing and Analysis

For the comparative elemental scans used to monitor surface composition of the CTO/LMO thin film superlattice, we collected lock-in signal intensity as a function of energy, N(E), with energy windows selected for each element we were interested in observing based on the location of their characteristic peaks.[35] The width of the energy window was minimized to only see the peak so as to decrease net collection time, while still capturing the shape of the peak as well as the peak-to-peak (P2P) maximum and minimum of the derivative of the lock-in signal (dN/dE). We performed 20 collection scans covering the energy range that contained the characteristic AES spectra for the elements of interest Ca, Ti, O, La, and Mn. The spectra were summed and their P2P values and numerically integrated areas under the curves (AUCs) were calculated with MATLAB scripts. The P2P and AUC values were then normalized to the oxygen P2P and AUC values, on the assumption that even if the overall signal intensity shifted from layer to layer (due to elemental brightness, surface roughness, or local field fluctuations driven by heater current changes) the effect would scale with the oxygen signal. The resulting values were then normalized to demonstrate overall elemental intensity shifts as a function of film thickness.

For the remainder of the elemental spectra, those shown are between 1 and 20 summed scans, smoothed and normalized to effectively demonstrate the observation of the characteristic peaks in comparison to one another. Likewise, the dN/dE plots have been shifted along the intensity axis so as to allow effective observation and comparison of all of the characteristic peaks' P2P signals.

## C. Thin Film Deposition

The CTO/LMO thin film superlattices were prepared via PLD using a 248 nm KrF excimer laser operating at a repetition rate of 1 to 5 Hz with dense polycrystalline targets prepared by solid state reaction. The laser spot size was varied from 2.5 to 6 mm$^2$, resulting in a net fluence of 0.8 to 2 J cm$^{-2}$. The chamber was evacuated to a base pressure below $5\times10^{-7}$ mbar before flowing high purity oxygen to create a background growth pressure of $5\times10^{-2}$ mbar. The single-crystal (110) NdGaO$_3$ substrate, purchased from CrysTec, was annealed in flowing O$_2$ for three hours at 1100°C prior to cleaning with acetone and then isopropyl alcohol in an ultrasonic bath. The substrate was heated to 750°C in growth pressure. Growth rate and thickness were observed with *in situ* RHEED using a 35 kV accelerating voltage. Thin films deposited for Auger spectra collection were grown using the same methods outlined above, with minor exceptions noted in the text. Auger signal as a function of background pressure studies were conducted on the same $NdGaO_3$ substrate as mentioned above with identical processing parameters. After heating to 750°C in growth pressure, the oxygen flow was stopped, and chamber pressure was allowed to drop to 5 x $10^{-7}$ mbar before the Auger scans were conducted. The pressure was then increased by increments of one order of magnitude and identical scans conducted, up to $10^{-1}$ mbar.

## III. RESULTS AND DISCUSSION

### A. Auger Probe Capabilities

The Staib AugerProbe[TM] has demonstrated its effectiveness for surface analysis in MBE systems with reported *in situ* observation of N, O, Si, Fe, Zn, Ga, Tb, and Dy.[32,33,36,37] However, to the best of our knowledge, *in situ* observation of complex oxides thin films grown by oxide PLD using this probe design has yet to be reported in the literature but other efforts on simple oxides have been reported.[32,36] We have, to date, observed characteristic Auger spectra for more than 24 elements using the probe, shown in Figure 2. The peaks observed range in energy from 44 to 1768 eV, and for many of the elements multiple transitions can be seen, including the fine structure. For example, shown in the insets of Figure 2, the primary oxygen $KL_{23}L_{23}$ peak is flanked at lower energies by the $KL_1L_{23}$ and $KL_1L_1$ peaks, respectively. All peaks are typically more easily observed when plotted as energy multiplied by intensity, or E*N(E), as opposed to N(E), but the dN/dE signal can reveal peaks which are otherwise subtle, especially those at low energy where there is a significant secondary electron background.

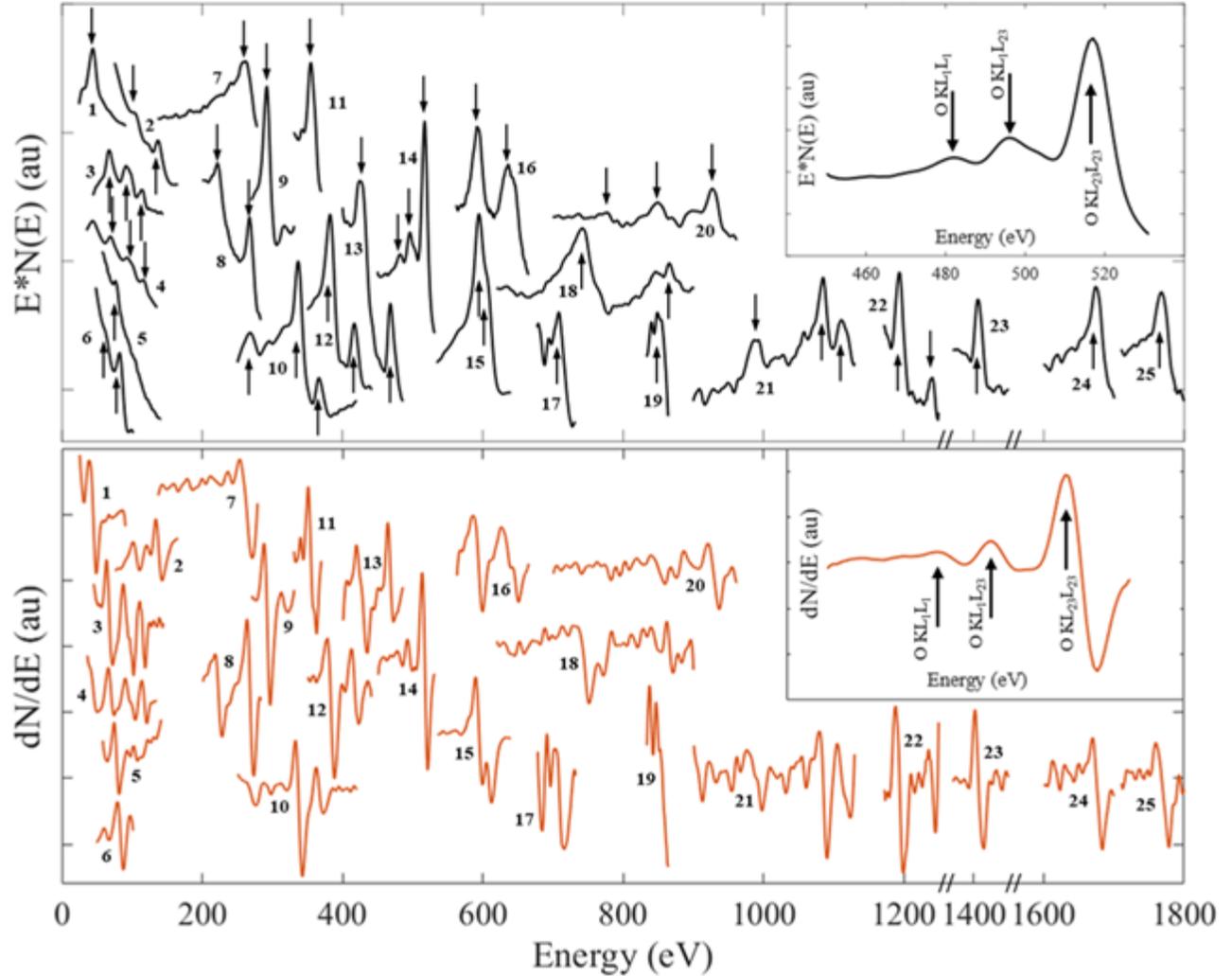

FIG. 2. Assorted spectra of 24 elements, intensity adjusted for visibility and comparison, plotted as E*N(E) (top) and dN/dE (bottom). Numbers correspond to spectra and peak identifications (for peaks marked with arrows) found in Table I.

The samples used to generate the spectra shown in Figure 2 were all either single crystal substrates purchased from CrysTec GmbH, thin films grown by PLD, or metal foils. The PLD targets used to grow the thin films were dense polycrystalline targets fabricated by solid state reaction except for the LaAlO$_3$ target, which is a single crystal target purchased from CrysTec GmbH. AES studies on metal foils were conducted at room temperature, while all other spectra were obtained at 750°C. Spectra 1, 22, and 23 were collected from a single crystal MgAl$_2$O$_4$ (111) substrate, 2 and 10 are from a GdScO$_3$ (110) substrate, 3 is from a CaY$_2$Co$_2$Ge$_3$O$_{12}$ thin film, 4 and 17 are from a Y$_3$Fe$_5$O$_{12}$ thin film, 5 is from a LaAlO$_3$ thin film, 6 and 19 are from a NdNiO$_3$ thin film, 7 is from the contamination on an Ag foil, 8 is from a SrRuO$_3$ thin film, 9, 12, and 14 are from a CaTiO$_3$ thin film, 11 is from an Ag foil, 13 is from a VO$_x$ thin film, 15 is from a BaTiO$_3$ thin film, 16 is from a LaMnO$_3$ thin film, 18 and 21 are from a NdGaO$_3$ (110) substrate, 20 is from a Cu foil, 24 is from a SrTiO$_3$ (100) substrate, and 25 is from a BaZrO$_3$ thin film. Many of the characteristic elemental peaks have been confirmed with multiple samples and sample-types; such as Ba and Ti from BaTiO$_3$ films, Ca and Zr from CaZrO$_3$ films, and Sr from SrTiO$_3$ thin films.

TABLE I. Identification of peaks denoted by arrows for the spectra shown and numbered in Figure 3, sorted by reference number (column 1). Column 2 is the energy at which the peak was observed, and column 3 is the element responsible for the peak, with questionable identifications noted. Column 4 is the atomic number, and column 5 is the identity of the transition responsible for the peak. The reference numbers marked with the ○ symbol are from single crystal substrates, the ↓ symbol indicates they are from PLD-grown thin films, and the § symbol indicates they are from metal foil.

| Ref. # | eV | Element | At. # | Transition |
|---|---|---|---|---|
| 1○ | 44 | Mg | 12 | LVV |
| 1 | 68 | Al | 13 | LVV |
| 2○ | 103 | Gd | 64 | NVV |
| 2 | 122 | Gd | 64 | NVV |
| 2 | 137 | Gd | 64 | NVV |
| 3↓ | 68 | Y | 39 | MNN |
| 3 | 91 | Ge (Y?) | 32 | MVV |
| 3 | 95 | Co | 27 | MVV |
| 3 | 112 | Ge | 32 | MVV |
| 4↓ | 43 | Fe | 26 | MVV |
| 4 | 68 | Y | 39 | MNN |
| 4 | 93 | Y | 39 | MNN |
| 4 | 116 | Y | 39 | MNN |
| 5↓ | 76 | La | 57 | NVV |
| 5 | 104 | La | 57 | NVV |
| 6↓ | 60 | Ni | 28 | MVV |
| 6 | 81 | Nd | 60 | NVV |
| 7§ | 260 | C | 6 | KLL |
| 8↓ | 222 | Ru | 44 | MNN |
| 8 | 267 | Ru | 44 | MNN |
| 9↓ | 292 | Ca | 20 | LMM |
| 9 | 316 | Ca | 20 | LMM |
| 10○ | 270 | Sc | 21 | LMM |
| 10 | 337 | Sc | 21 | LMM |
| 10 | 367 | Sc | 21 | LMM |
| 11§ | 354 | Ag | 47 | MNN |
| 12↓ | 382 | Ti | 22 | LMM |
| 12 | 416 | Ti | 22 | LMM |
| 13↓ | 427 | V | 23 | LMM |
| 13 | 468 | V | 23 | LMM |
| 14↓ | 481 | O | 8 | KLL |
| 14 | 496 | O | 8 | KLL |
| 14 | 517 | O | 8 | KLL |
| 15↓ | 594 | Ba | 56 | MNN |
| 15 | 604 | Ba | 56 | MNN |
| 16↓ | 592 | Mn | 25 | LMM |
| 16 | 636 | La | 57 | MNN |
| 16 | 643 | Mn | 25 | LMM |
| 17↓ | 708 | Fe | 26 | LMM |
| 18○ | 742 | Nd | 60 | MNN |
| 18 | 847 | Ga | 31 | LMM |
| 18 | 865 | Nd | 60 | MNN |
| 19↓ | 848 | Ni | 28 | LMM |
| 20§ | 777 | Cu | 29 | LMM |
| 20 | 849 | Cu | 29 | LMM |
| 20 | 927 | Cu | 29 | LMM |
| 21○ | 987 | Ga | 31 | LMM |
| 21 | 1085 | Ga | 31 | LMM |
| 21 | 1111 | Ga | 31 | LMM |
| 22○ | 1194 | Mg | 12 | KLL |
| 22 | 1241 | Mg | 12 | KLL |
| 23○ | 1406 | Al | 13 | KLL |
| 24○ | 1674 | Sr | 38 | LMM |
| 25↓ | 1768 | Zr | 40 | LMM |

Our ultimate objective is to collect Auger spectra in real time during deposition, with the ability to perform quantitative or semi-quantitative analysis. Real time spectra acquisition will broaden the scope of our analytical capabilities and potentially answer important fundamental questions about the precise dynamics of surface evolution of complex oxides during pulsed laser deposition. Likewise, fully harnessing the sensitivity of this probe for quantitative results, as demonstrated during MBE growth,[33] will provide a powerful lens to identify subtle shifts in stoichiometry and surface termination of deposited films. The primary obstacles to realizing quantitative real-time *in situ* Auger analysis are the development of standards and optimization of the growth system for both Auger signal and deposition parameters, which will be addressed here.

Developing standards for quantification of the stoichiometry of complex oxides grown with PLD is a distinct challenge due to the common deposition of multi-species materials, which require a more complex accounting of relative signal ratios for compositional comparisons. Additionally, there can be unique relationships between PLD growth conditions and the resulting stoichiometric composition of the thin films, depending on the materials deposited.[38] However, despite the challenge of developing a robust quantitative analysis method, the wealth of information *in situ* stoichiometry can provide regarding the composition-property relationships of complex oxide thin films and heterostructures is profound.[13,14,18,38]

Developing growth chamber parameters for simultaneous Auger signal acquisition and thin film deposition is, primarily, a matter of optimization. However, two obstacles addressed by the design of the Staib AugerProbe™ are the high-pressure requirements for typical complex oxide growth with PLD, and the real-time requirement of scan acquisition timescales comparable to the growth rate of the material being observed. To identify the acceptable range of growth pressures which can be used with the *in situ* probe, we conducted spectral acquisitions as a function of background pressure on a $NdGaO_3$ (110) substrate prepared in the same manner as those used for typical growths. As can be seen in Figure 3, the signal only decreases by approximately 10% up from $5\times10^{-7}$ to $5\times10^{-4}$ mbar, and 50% at $5\times10^{-3}$ mbar, with effectively complete loss at $5\times10^{-2}$ mbar. The oxygen peaks, each 20 summed scans, is shown in the inset of Figure 3, and is clearly still present even at $5\times10^{-3}$ mbar. Therefore, adequate signal for analysis should be attainable with most deposition conditions used for complex oxides.

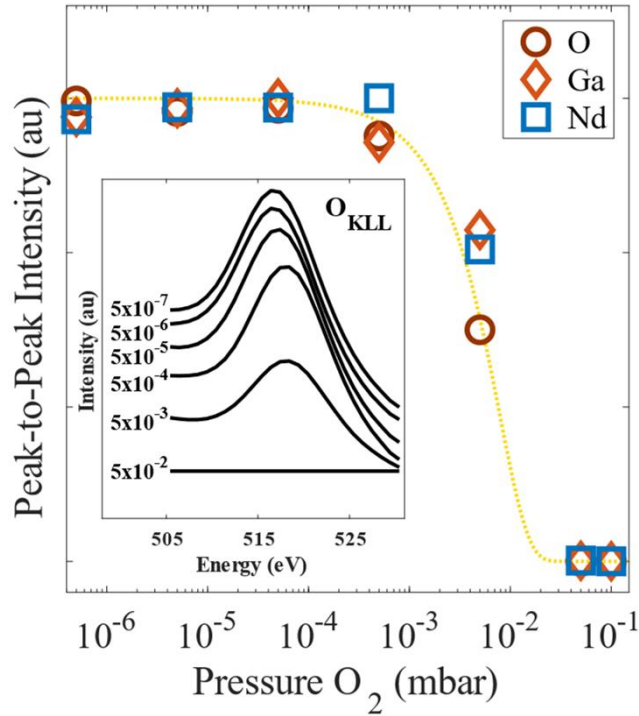

FIG. 3. Normalized O, Ga, and Nd peak-to-peak intensities of 20 summed scans (per element) as a function of chamber pressure, and (inset) 20 summed oxygen KLL peak spectra with denoted pressure, intensity shifted linearly for clarity. Approximately 50% of the oxygen Auger signal is lost at $5 \times 10^{-3}$ mbar, and effectively all of it lost by $5 \times 10^{-2}$ mbar. The dashed line is a guide to the eye.

## B. *Depth-Dependent Surface Composition Analysis*

We optimized PLD growth parameters for the CTO/LMO superlattice to achieve long-lived layer-by-layer growth with a smooth resulting surface as observed with RHEED (shown in Figure 4). Deposition of CTO causes a very sharp increase in specular spot intensity, while the deposition of LMO causes a rapid decay in intensity. This results in a step-like RHEED pattern, with the tops of the steps corresponding to the deposition of CTO, and the bottoms corresponding to LMO deposition. Intensity oscillations corresponding to the growth of a single monolayer are clear for the deposition of both materials, following the larger intensity trends mentioned above. This clearly indicates that the superlattice grown consists of alternating layers of CTO and LMO, where each layer consists of four atomic layers of the complex oxide material, as shown in Figure 4(a). The quality of the film during the deposition process was observed by monitoring the electron diffraction pattern. At the end of the growth the capping layers of the superlattice were LMO, and the post-growth diffraction pattern demonstrates this in Figure 4(b). The post-growth film was smooth, as indicated by the vertical streaking of the diffraction pattern due to Bragg columns arising from the two-dimensional surface. The mid-growth diffraction patterns of the CTO surface (Figure 4(c)) and LMO surface (Figure 4(d)) show equivalent smoothness and no observable three-dimensionality, as would be indicated by a grid of diffraction spots. Therefore, this is a suitable system for AES surface composition studies. The stability of the layer-by-layer growth allows us to use the same growth parameters used for the superlattice, with RHEED, to perform surface composition analysis with single-atomic-layer resolution.

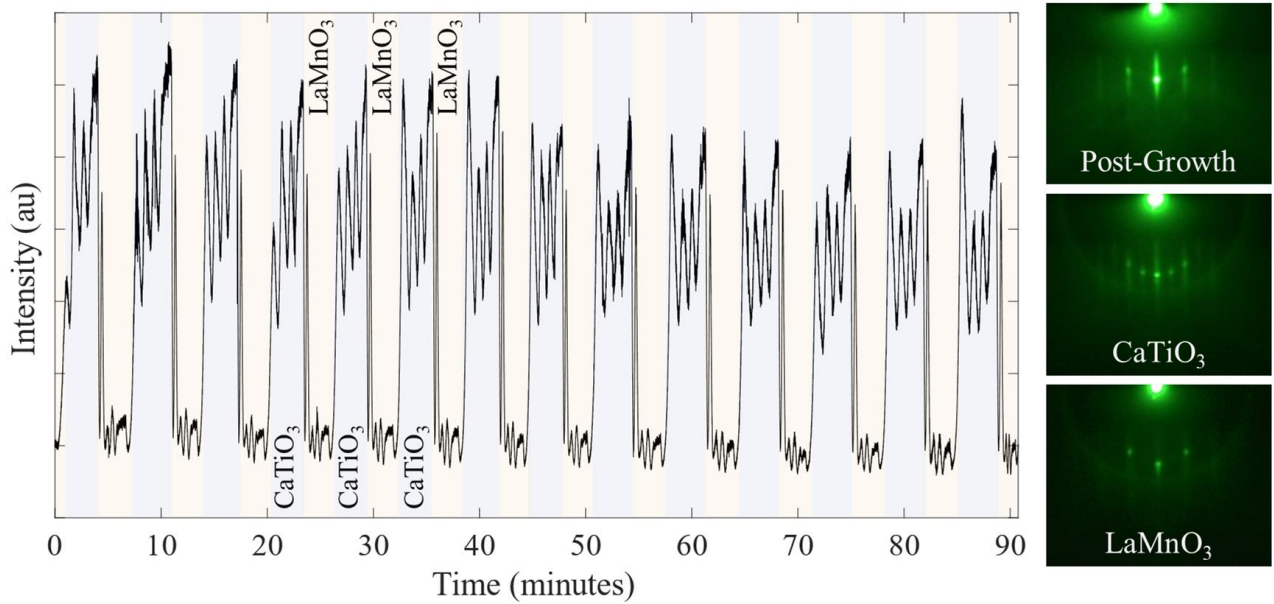

FIG. 4. Electron diffraction data from the deposition of a $CaTiO_3$/$LaMnO_3$ superlattice on a $NdGaO_3$ single crystal substrate. (a) RHEED oscillations during growth, with four oscillations per material, per layer, and distinction between the two materials indicated by denoted background color. (b-d) Electron diffraction patterns demonstrating smoothness of the as-grown film after growth (b), after deposition of four atomic layers of CTO (c), and after deposition of four atomic layers of LMO (d).

Surface composition of the superlattice, acquired with *in situ* AES, demonstrates that, as expected, the elemental signals from a deposited material increase with the deposition of atomic layers and saturate after a few (approximately five) monolayers, as shown in Figure 5. Likewise, the elemental signal from the underlayer decreases with the deposition of an alternate material on top and vanishes after approximately five monolayers. With a lattice constant of approximately 4 Å, this corresponds to a depth limit of around 2 nm for the Auger probe in this geometry. The large increase in signal observed after the deposition of a single monolayer is more than adequate for measurement, demonstrating that the probe is capable of sub-monolayer resolution. The ratio of A-site to B-site (Ca/Ti or La/Mn) signal is consistent throughout the depositions, indicating that the data acquisition and analysis methods used are suitable for this degree of qualitative analysis, and, with the addition of standards, may be suitable for quantitative analysis.

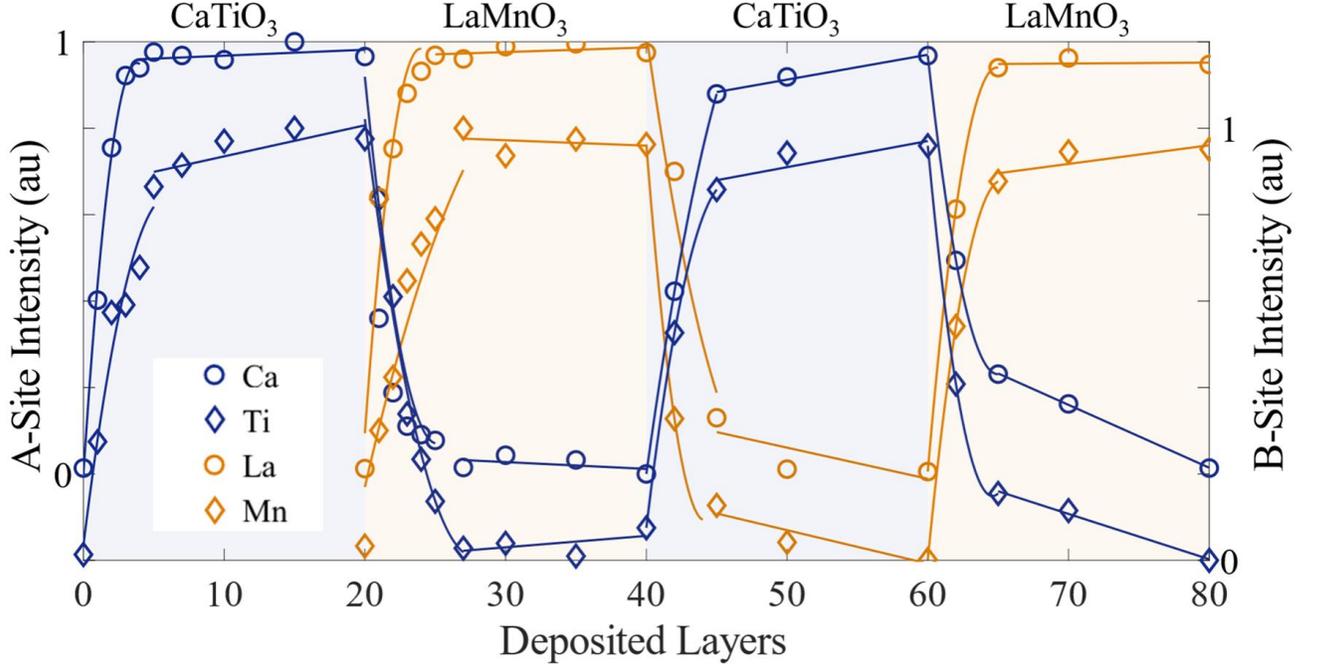

FIG. 5. A plot of normalized (to the oxygen signal) elemental Auger spectra signals during the growth of $CaTiO_3$/$LaMnO_3$ superlattices. Note that the A-site intensity (Ca and La) data is offset from the B-site intensity (Ti and Mn) data for clarity. Each data point is from an area under the curve calculated from twenty summed scans of the characteristic peak for that specific element normalized to the oxygen signal from that specific collection. The Auger spectra were collected *in situ* between depositions of known thickness, as calibrated by RHEED and previous growths. Lines are guides to the eye.

The sensitivity of the Auger probe to a given material is dependent on the ability to generate and observe Auger electrons. Generation is predominantly a function of the amount of the material present, though other factors such as ionization cross section play a role. Observation of the Auger electrons is then dependent on their ability to escape the material and find their way to the detector. The escape process is largely responsible for surface-sensitivity, which is why AES is a valuable surface characterization tool, because the relatively low energy of Auger electrons limits their ability to leave the parent material. The proportion of Auger electrons which escape can be approximated as

$$\frac{N}{N_0} = e^{\left(\frac{-z}{\lambda}\right)} \qquad (1)$$

where z is the depth from which they are generated and λ is the inelastic mean free path ($\lambda_{MFP}$). However, the $\lambda_{MFP}$ for any given element is not a constant and depends on the other elements present as well as the structure of the material, as there are numerous energy-dependent processes which can prevent an Auger electron from escaping. For this reason, experimental escape depth data was fit by Seah and Dench[39] and the following relation was found:

$$\lambda_{MFP} = \frac{143}{E^2} + 0.054(E)^{1/2} \quad 2$$

where E is the energy of the Auger electron in eV, and $\lambda_{MFP}$ is in nm.

The deposition data from the $CaTiO_3$/$LaMnO_3$ superlattice growth (Fig. 5) was fit to the basic escape model (Eq. 1), with the results shown in Figure 6. The fits were applied assuming a single layer thickness of 0.4 nm using Auger energies shown in Table II. The escape depths calculated from the fit are quite close to the theoretical values and follow the expected trend. There are many reasons the observed values may vary from expected values, including the structure or composition of the parent material or the material deposited on it. These results show this probe has the acute surface sensitivity expected for AES, making it a viable tool for delicate real time *in situ* surface characterization experiments.

TABLE II. Escape depths calculated from fits made to deposition data shown in Fig. (5), following Eq. (1), compared to escape depths calculated for the same elements following Eq. (2).

| Element | Transition | Energy (eV) | Fit $\lambda_{MFP}$ (nm) | Calculated $\lambda_{MFP}$ (nm) |
|---|---|---|---|---|
| Ca | LMM | 291 | 0.60 | 0.92 |
| Ti | LMM | 387 | 1.20 | 1.06 |
| La | NVV | 80 | 0.51 | 0.51 |
| Mn | LMM | 590 | 1.23 | 1.31 |

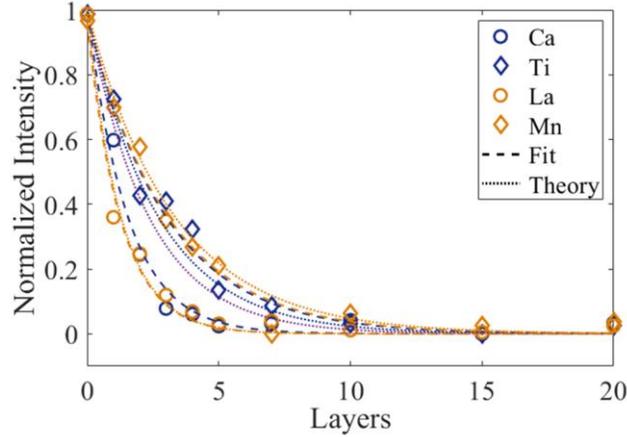

FIG. 6. Fits to the normalized intensity data of $CaTiO_3$/$LaMnO_3$ superlattice layers during deposition. Comparing the fits (long dash) to the calculated escape depths (short dash) for these elements reveals that the escape depth is proportional to the Auger electron energy, as expected, and has agreement within one unit cell (0.4 nm) for all elements.

## IV. CONCLUSION

In summary, we have demonstrated *in situ* Auger electron spectroscopy of complex oxide thin film surfaces grown by PLD. Characteristic Auger spectra have been collected *in situ* from 24 elements sourced from various complex oxide single crystal substrates, PLD-grown thin films, and metal foils, displaying the viability of this technique for a wide variety of materials analysis applications in our growth chamber. Challenges limiting our ability to quantify our results have been identified and discussed, and strategies have been proposed to overcome them moving forward. Sustained layer-by-layer growth of CTO/LMO superlattices on $NdGaO_3$ substrates was achieved, thereby allowing monolayer depth-resolved Auger surface composition studies. We have found that our Auger probe provides spectra from, approximately, the top 2 nm of the thin film, with the possibility of sub-monolayer depth resolution. The prospect of applying Auger electron surface analysis in real-time during pulsed laser deposition of complex oxides has been addressed.

## ACKNOWLEDGMENTS

The authors gratefully acknowledge support from Staib Instruments, as well as the benefit of discussions with Dr. Philippe G. Staib, Dr. Eric Dombrowski, and Laws Calley. This work was supported by the Air Force Office of Scientific Research under contract FA9550-16-1-0335. T.O and M.S acknowledge the Andrew and Erna Viterbi Graduate Student Fellowship.